# High resolution x-ray study of nematic-smectic-A and Smectic-A-reentrant nematic transitions in liquid crystal+aerosil gels

M. Ramazanoglu[1,4], S. Larochelle[1], C.W. Garland[2] and R.J. Birgeneau[1,3]

[1]Department of Physics, University of Toronto, Ontario M5S 1A7, Canada

[2]Department of Chemistry, Massachusetts Institute of Technology, Cambridge, MA 02139

[3]Department of Physics and Lawrence Berkeley Laboratory, University of California,

Berkeley, CA 94720

[4]Department of Physics & Astronomy, McMaster University, Ontario L8S 4M1, Canada

(Dated: November 5, 2007)

We have studied the effects of quenched random disorder created by dispersed aerosil nanoparticle gels on the nematic to smectic-A (N-SmA) and smectic-A to reentrant nematic (SmA- RN) phase transitions of  thermotropic liquid crystal  mixtures of 6OCB (hexyloxycyanobiphenyl) and 8OCB (octyloxycyanobiphenyl).  These effects are probed using high-resolution synchrotron x-ray diffraction techniques.  We find that the reentrant characteristics of the system are largely unchanged by the presence of the aerosil gel network.  By comparing measurements of the smectic static structure amplitude for this 8OCB-6OCB+aerosil system with those for butyloxybenzilidene-octylaniline (4O.8)+aerosil gels, we find that the short-range smectic order in the smectic-A phase is significantly weaker in the reentrant system.  This result is consistent with the behavior seen in pure 8OCB-6OCB mixtures.  The strength of the smectic ordering decreases progressively as the 6OCB concentration is increased.  Detailed line shape analysis shows that the high- and low-temperature nematic phases (N and RN) are similar to each other.

[2] Electronic address: cgarland@mit.edu

[3] Electronic address: chancellor@berkeley.edu

[4] Electronic address: mehmet@physics.mcmaster.ca

## I. INTRODUCTION

Recently there have been a number of studies of the phase transition behavior associated with the smectic-A (SmA) phase in gels formed by dispersing aerosil particles in a liquid crystals (LC) – to be denoted LC+aerosil gels. The purpose of these measurements has been to probe the effects of the quenched random disorder created by the gel formed by the silica nanoparticles on the second-order nematic–smectic-A (N-SmA) and smectic-A–smectic-C (SmA-SmC) phase transitions and also on the first-order phase transition from isotropic to smectic-A (I-SmA) [1-6]. Thermotropic LCs are ideal materials for studying the effects of quenched randomness since they show a rich variety of phases as a function of temperature. Further, it is generally possible to prepare homogeneous samples. Importantly, the pure LC phases are typically well characterized not only by x-ray scattering but also by calorimetry, light scattering, and dielectric studies, especially for commonly studied systems such as 8OCB, 8CB, and 4O.8 [7-9]. Generally, the range from weak quenched disorder ($\rho_S$ =0.025 g of $SiO_2$/$cm^3$ of LC) to strong disorder ($\rho_S$ =0.3) can be accessed with LC+aerosil gel systems with minimal uncertainty in the concentrations. Typically most of the interest has been focused on the range where the disorder is weak.

In the present high-resolution x-ray study, we focus on nematic reentrance in dispersed LC+aerosils. The reentrant nematic phase (RN) is one of the interesting characteristics of some LCs. It may be obtained when an appropriate mixture is made of two similar liquid crystals, one with and one without a smectic-A phase. In all cases to date, one or both of the liquid crystal molecules are polar. In the present case, these two LCs are the alkyloxycyanobiphenyl homologs 8OCB and 6OCB. Reentrant behavior in LCs was discovered by Cladis in a mixture of HBAB (hexyloxybenzylidene-aminobenzonitrile) and CBOOA (cyanobenzylidene-octyloxyaniline) [10]. The RN phase can also be realized by increasing the pressure on a single LC [11]. The polar liquid crystal 8OCB has an incommensurate partial bilayer smectic-$A_d$ (Sm$A_d$) phase, while its homolog 6OCB shows a N phase but no Sm$A_d$ phase. The phase sequence and transition temperatures for these LCs are for 8OCB [12] and 6OCB [13], respectively. In mixtures of 8OCB and 6OCB LCs, to be denoted [8:6]OCB, the phase sequence on cooling is isotropic → nematic → smectic-$A_d$ → reentrant nematic. The reentrant nematic phase typically occurs on cooling just before the LC crystallizes.

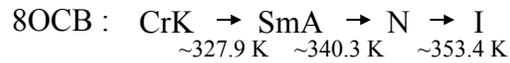

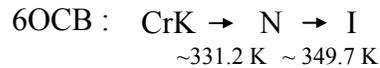

Kortan *et al*. [14] studied [8:6]OCB mixtures using high-resolution x-ray scattering techniques. According to their study, the SmA phase boundary shown in Fig. 1 is parabolic in the temperature – concentration ($T$-$y$) plane. Here, $y$ =[6OCB]/[8OCB] is the mole ratio of 6OCB to 8OCB. The composition can also be expressed as a mass ratio by dividing $y$ by 1.10. The median temperature $T_m$, which is defined as the midpoint of the temperature range for the smectic-A phase, was found to be 311.2 K [14]. The "nose" of the parabolic SmA phase boundary corresponds to $y_o = 0.427$; that is, for $y > 0.427$ there is no SmA phase in the pure LC mixtures. Concomitantly, the strength of the SmA ordering decreases to zero as $y \rightarrow y_o$ at $T = T_m$. In most previous work on LC+aerosil systems, quenched random disorder effects were studied in LC samples where the smectic phases are fully developed [1-4, 6]. As found in these experiments, even the weakest disorder imposed by the aerosil gel, $\rho_S = 0.025$ g/cm$^3$, modifies the pure N-SmA transition properties dramatically. The quasi long range order (QLRO) of the smectic phase in a pure LC is destroyed by the aerosil gel, and only short range order (SRO) persists. In the present [8:6]OCB+aerosil study, the same effects are investigated for a smectic phase which is weakly ordered in the pure LC system, especially close to the instability limit $y_o$ [14].

This study is largely empirical. Specifically, we have investigated whether or not the reentrant characteristics of a pure LC mixture are preserved in the presence of quenched random disorder over a range of disorder strengths.

## II. EXPERIMENTAL TECHNIQUES

8OCB and 6OCB LC samples were purchased from Frinton Laboratories [15], and hydrophilic type-300 aerosil silica material was obtained from Degussa Corp. [16]. Both the LC materials and the silica aerosil were used in these experiments without any further purification.

X-ray diffraction experiments were carried out on beam line X-22-A at the National Synchrotron Light Source of the Brookhaven National Laboratory and on beam line B-2-1 at the Stanford Synchrotron Radiation Laboratory. These are both bent magnet beam lines with Si(111) monochromators. The analyzer crystal used in this study was also a perfect Si(111) crystal. The initial x-ray beam energy was tuned to 10.7 keV for both beam lines. A detailed description of the sample preparation and x-ray scattering techniques can be found in recent publications [1].

## III. RESULTS AND ANALYSIS

In previous LC+aerosil experiments [1-3,6], a line shape analysis protocol has been developed to distinguish the scattering due to the pretransitional thermal fluctuations and that due to the developing smectic layers. The details of the latest version of this model can be found elsewhere [1]. The total cross-section, which consists of thermal and static contributions, is the spherical average of a presumed microscopic structure factor given by

$$S(\boldsymbol{q}) = S^{Thermal}(\boldsymbol{q}) + S^{Static}(\boldsymbol{q}) \qquad (1)$$

where

$$S^{Thermal}(\boldsymbol{q}) = \frac{\sigma_1}{1 + (q_\parallel - q_o)^2 \xi_\parallel^2 + q_\perp^2 \xi_\perp^2 + c q_\perp^4 \xi_\perp^4} \qquad (2)$$

and

$$S^{Static}(\boldsymbol{q}) = \frac{a_2 (\xi_{\parallel 2} \xi_{\perp 2}^2)}{[1 + (q_\parallel - q_o)^2 \xi_{\parallel 2}^2 + q_\perp^2 \xi_{\perp 2}^2]^2} \qquad (3)$$

Here, $q_\parallel (q_\perp)$ is the parallel (perpendicular) momentum component of the smectic wave vector $\boldsymbol{q_o} = 2\pi / d \, \mathbf{Z}$, where $d$ is the SmA layer spacing, $\mathbf{Z}$ is the unit vector along the nematic director, and $\xi_\parallel (\xi_\perp)$ is the parallel (perpendicular) correlation length for the thermal structure function. The parallel correlation length $\xi_{\parallel 2}$ in the static structure factor is taken to be a global (i.e., temperature independent) variable, while the thermal parallel correlation length is treated as a temperature-dependent free parameter. During the fitting analysis, the fourth-order correction amplitude $c$ and the perpendicular correlation length are calculated assuming the same behavior relative to the parallel correlation length as in the pure LC mixtures [14]. Since the static peak profile is insensitive to a fourth-order correction term, this term has been excluded from Eq. (3). The direct beam profiles were collected after each temperature scan to define the resolution function individually. This profile was then convoluted numerically with the spherical average of Eq. (1), and the resultant function was fit to the scattering data. The background, due primarily to the aerosil particles, was assumed to have the form $b_c + b_P/q^4$, where $b_P$ is the Porod amplitude and $b_c$ is a constant. The analysis was carried out over a typical $q$ range of [0.1,0.35]; this range excludes the signal from the Kapton cell window.

A representative set of fits to the scattering intensity with our smectic structure factor in [8:6]OCB+aerosil gels is shown in Fig. 2 for the $y = 0.405$, $\rho_S = 0.025$ sample at a number of temperatures which cover temperature region from N to RN phases. The least-squares fitting is carried out for all temperature scans simultaneously with four free parameters that are temperature dependent ($\xi_{\parallel}$, $\sigma_1$, $a_2$, $q_o$) and three global parameters that are independent of $T$ ($\xi_{\parallel 2}$, $b_P$, $b_c$). We will call this a *global* fit. The solid lines are the results of such a fit with our model for the structure factor, namely the spherical average of Eq. (1), (2) and (3). Clearly, the global model describes the x-ray scattering data quite well. One can, however, see subtle deviations in the wings, especially at low temperatures. To explore a more flexible fitting form, we allowed the background to be temperature dependent. Such fits will be called *free* since there are five free *T*-dependent parameters and only the static parameter $\xi_{\parallel 2}$ is assumed to be independent of temperature. The bottom two panels of Fig. 2 show free fits for data at the same temperatures as the global fits given in panels *g* and *h*.

As seen from Fig. 2, the scattering peaks are asymmetric at all temperatures; this is a characteristic of the spherical average of data from an anisotropic microscopic structure factor. The amplitude of the static part of the structure factor, $S^{Static}$, starts to increase at ~317 K and reaches its maximum at ~311 K. This latter temperature corresponds to the midpoint between the two smectic boundaries, as found previously in the study by Kortan *et al.* [14] of the pure LC mixtures. As may be seen from the inset on panel *d*, $S^{Static}$ dominates $S^{Thermal}$ near $q_o$, but it never reaches the point where $S^{Static}$ alone is enough to define the entire SmA peak. In contrast to this, for butyloxybenzilidene-octylaniline (4O.8), the 4O.8+aerosil system deep in the SmA phase shows that $S^{Thermal}$ contributes only a weak broad background while $S^{Static}$ alone defines almost all of the scattering peak [1]. Therefore, we can empirically conclude that the SmA phase in a [8:6]OCB+aerosil gel is formed very weakly. This conclusion is supported by the normalized maximum amplitude $a_2$ values listed in Table I, which will be discussed in some detail below.

As may be seen in Fig. 2, on further cooling below 311 K the total x-ray intensity gradually decreases and concomitantly the relative contribution of $S^{Thermal}$ to the total intensity increases. Finally, at low temperatures, $S^{Thermal}$ dominates the cross section, as it does in the high-temperature nematic phase. For example, in the $y = 0.405$, $\rho_S = 0.025$ sample at $T = 298.65$ K, the $S^{Static}$ term has completely vanished. At this temperature, the sample is in the low-temperature reentrant nematic (RN) phase. On cooling, the static

amplitude $a_2(T)$ behaves qualitatively like the static structure amplitude in the

4O.8+aerosil system [1] for temperatures down to $T_m$. The value of $a_2(T)$ becomes

nonzero for $T < T_1$*, and it reaches a maximum value $a_2(T_m)$ at $T = T_m$ where the $S^{Static}$

term dominates over $S^{Thermal}$ at the peak position $q_0$. It then decreases on further cooling and becomes zero for $T < T_2^*$. This can be seen from Figs. 2- 4. The systematic variation of $a_2(T)$ is shown for several LC concentrations $y$ and various $\rho_S$ values in Fig. 3. A clear view of the effect of $y$ on $a_2(T)$ is provided by the $\rho_S = 0.025$ gel data shown in Fig. 4.

The normalized maximum static amplitude values, $a_2(T_m)/b_P$, are also listed in Table I.

The $a_2$ values in Figs. 3 and 4 come from global fits, but the behaviors of all the fitting parameters are essentially the same for both the global and the free fits. The one exception is the behavior of $q_0(T)$ in the RN phase, which will be discussed at the end of this section. The parameter values presented for $\xi_\parallel$, $\sigma_1$, $\xi_{\parallel 2}$, and $a_2$ will all be those derived from the global fits.

As can be seen from the reentrant $a_2$ values when they are compared to

corresponding $a_2$ values for 4O.8+aerosils [Tables I and II in Ref. 1], the SmA short-

range order is quite weak in the [8:6]OCB+ aerosil systems. For example, the maximum

value of $a_2/b_P$ for the $y = 0.33$, $\rho_S$ =0.025 sample, which is the largest such value in Table

I, is comparable to that of a 4O.8+aerosil gel sample with $\rho_S$ =0.062. In other words, the least disordered [8:6]OCB+aerosil gel with $y = 0.33$ compares to a 4O.8+aerosil gel with substantial disorder (larger $\rho_S$). All other maximum $a_2/b_P$ values are found to become progressively smaller with increasing $y$ and $\rho_S$, decreasing to as little as 1% of the values observed in non-reentrant systems. This result is consistent with the assumptions of various theoretical studies [17-19]. On further cooling below $T_m$, $a_2(T)$ decreases and finally below a certain temperature, the $S^{Static}$ term completely vanishes (that is, $a_2 = 0$). This temperature is the effective SmA-RN transition temperature $T_2$*. The temperature values $T_1$* and $T_2$*, which are the effective transition temperatures for N-SmA and SmA-RN, are listed in Table I. In the 4O.8+aerosil study [1], when the "low temperature" $a_2$ values ($a_{2LT}$, obtained ~7K below the N-SmA transition temperature) are normalized by the Porod amplitude to account for the volume of sample in the x-ray beam, an empirical power-law behavior $a_{2LT}/b_P \sim \rho_S^{-1}$ was found to hold. This demonstrated that the scattering intensity was proportional to the LC mass in the beam. As shown in Fig. 5(a), the same scaling behavior is not observed in the [8:6]OCB+aerosil gel system. This reflects directly the weak smectic order in [8:6]OCB+aerosil gels. In panel b of Fig. 5, the smectic order [as given by the normalized static amplitude values $a_2(T_m)/b_P$] is shown as a function of the composition variable $y$. The smectic order rapidly decreases as the value of $y$ increases, as one would expect, although the $a_2$ values for the $y = 0.405$ and $y = 0.42$ samples are somewhat irregular. We presume that this originates from a slight error (~0.01 to 0.02) in y for one or both samples.

The temperature range investigated in the current experiment is ~55 K, roughly twice that used in any previous LC+aerosil study [1-4,6]. Since there are two phase transitions separated by ~30 K to ~10 K depending on the $y$ value, the temperature scans need to cover the range from ~345 K to ~290 K, whereas, for example, the 4O.8+aerosil gel measurements range from 340 K to 320 K [1]. Keeping in mind that a complete scan for a gel sample is restricted to two hours of x-ray illuminance in order to avoid possible x-ray damage [1], data cannot be acquired at very small temperature increments for a given scan as could be done in previous LC+aerosil studies [1-4]. This more limited data set greatly diminishes our ability to carry out a power-law analysis of $a_2(T)$ values [ $a_2 \sim \left| T - T^* \right|^{2\beta}$ ], as was done in Refs. [1-3] and [6]. Thus, values for the critical exponent β are reported in Table I only for the lowest density sil concentration where the SmA range is largest. Some qualitative comments about the exponent $\beta$ will be given in Sec. IV.

Throughout the analysis of [8:6]OCB+aerosil data, the static parallel correlation lengths $\xi_{\parallel 2}$ were always treated as temperature-independent fitting parameters. The best least-squares $\xi_{\parallel 2}$ values are given in Table I, and these correlation lengths decrease with increasing ρ$_S$ as one would expect. The variation in $\xi_{\parallel 2}$ as a function of $\rho_S$ for several values of $y$ is shown in Fig. 6($a$). There is also a systematic decrease in the static correlation lengths as a function of the concentration $y$, as shown in Fig. 6($b$). This variation $\xi_{\parallel 2}(y)$ is especially strong for the $\rho_S$ =0.025 samples. This is the expected behavior since as the 6OCB concentration increases, one is moving progressively toward the "nose", beyond which there is only a nematic phase. Surprisingly, for the other two gel densities ($\rho_S$ =0.062 and 0.15), $\xi_{\parallel 2}$ was found to vary rather slowly with $y$, and the scatter of the sparse data makes it difficult to characterize the trend of $\xi_{\parallel 2}$ with $y$. The proposed line shape model used above is based on a static term with only one length, which is the temperature-independent parallel correlation length. We also tried analyzing the measured profiles by allowing the perpendicular correlation length $\xi_{\perp 2}$ for the static structure factor term to vary freely as a global fitting parameter. However, $\xi_{\perp 2}$ values were found to be indeterminate in such fits as might be expected since after the spherical averaging of Eq. (1), $\xi_{\perp 2}$ has only a secondary effect on the line shape.

In the current analysis, the thermal structure term is allowed to be a function of temperature as in previous studies. The fit values for the thermal correlation length $\xi_{\parallel}$ and thermal amplitude $\sigma_1$ are shown in Fig. 7 and Fig. 8, respectively. First, let us consider the two samples with $y = 0.46$ and $y = 0.50$ that do not have any SmA phase. As shown in Fig. 7($a$) and Fig. 8($a$), $\xi_{\parallel}$ and $\sigma_1$ both exhibit maxima at ~310 K, which is close to $T_m$ for all the other samples, and the peak values for $y = 0.50$ are much smaller than those for $y = 0.46$. Thus strong SmA fluctuations exist in the nematic phase near the nose of the transition line, but such fluctuations die off rapidly as $y$ increases beyond $y_o$.

Now consider the samples with $y \leq y_o$ where the phase sequence N − SmA − RN is observed; see Figs. 7($b$-$d$) and 8($b$-$d$). On cooling from high temperatures, the smectic fluctuations grow and both the thermal correlation length $\xi_{\parallel}$ and thermal amplitude $\sigma_1$ increase rapidly up to peak values at $T_1^*$, the effective N-SmA transition temperature. Further cooling below $T_1^*$ leads to a rapid decrease in both $\xi_{\parallel}$ and $\sigma_1$ together with a concomitant growth in $a_2$. This behavior is just like that for the usual N-SmA transition in a LC+aerosil system [1] and represents the increasing amplitude of the smectic layer ordering. However, the growth of SmA order and the concomitant decrease in thermal fluctuations saturates at the midpoint temperature $T_m$. On further cooling below $T_m$, the smectic order begins to break down and the thermal fluctuations begin to increase again. Both $\xi_{\parallel}$ and $\sigma_1$ exhibit second sharp maxima at $T_2^*$, which is the lower SmA-RN phase transition temperature. At temperatures below $T_2^*$, the static amplitude $a_2$ becomes zero as the smectic-A phase "melts" into the reentrant nematic phase. The $T_1$* and $T_2$* values obtained from the position of the $\xi_\parallel$ and $\sigma_1$ peaks agree within their error bars with the values obtained from $a_2(T)$. As is evident in Figs. 6 and 7, the values for $\xi_\parallel$ and $\sigma_1$ in the RN phase are similar to those in the N phase.

It is interesting to note that, by definition, data analogous to those shown in Fig. 7 and 8 do not exist in the reentrant systems which are "pure", that is, do not have the aerosil gel network. Rather, in the pure systems, the structure factor in the N and RN phase is a modified Lorentzian whereas in the smectic phase the structure factor is a power law singularity. In both the nematic and smectic phases the scattering is entirely dynamic. By contrast, in the measurements of the liquid crystal mixtures embedded in the aerosil gel network, there are distinct static and dynamic (thermal) components and the data analysis allows one to separate clearly the two components. Specifically, both the thermal correlation length and the susceptibility exhibit power law maxima at the N-SmA and the SmA-RN transition. The maximum thermal length is bounded by the corresponding temperature independent static correlation length.

One uncertain feature in fitting the $I(q)$ data on the present reentrant LC + aerosil system is the behavior of the background. There is strong evidence from measurements on several liquid crystals in the N phase far above $T_{NA}$ that the form $b_c + b_P/q^4$ is a very good approximation for the background scattering [1,3]. Since $b_P$ is due to the incoherent aerosil scattering while $b_c$ is a small term due mostly to the cell windows and the air scattering, one would expect both $b_P$ and $b_c$ to be independent of $T$, as is assumed in our global fits. However, the resulting $q_o(T)$ behavior in the RN phase is physically unattractive, as described below. Thus we also carried out free fits in which $b_P$ and $b_c$ were allowed to vary with $T$.  The resulting background fitting parameters are shown for several concentrations and different gel densities in Fig. 9.  In each case, $b_P$ and $b_c$ are found to decrease with decreasing temperature.  The global values of $b_P$ and $b_c$ are also shown in Fig. 9.  Good agreement between the free and global values in the SmA region is not surprising since $I(q)$ is very large there, and SmA data will dominate the background aspects of the global least-squares fitting procedure.  We have no physical explanation for a $T$-dependent background, but it is important to emphasize that such a background variation does not significantly alter any important LC+aerosil properties determined in our analysis.  A fitting analysis in which both $b_P$ and $b_c$ are taken to be temperature-dependent parameters (free fits) yields $\xi_\parallel(T)$, $\sigma_1(T)$ and $a_2(T)$ behavior that is essentially the same as that shown in Figs. 3, 4, 7 and 8, and the static $\xi_{\parallel 2}$ values from free fits are essentially the same as those given in Table I.

As stated earlier, the behavior of the pure LC mixtures was used to characterize the thermal perpendicular correlation lengths and the fourth-order correction coefficient $c$, assuming that the pure LC relative critical dependencies held for the LC+aerosil gel samples. In addition, some secondary fits were carried out using different assumptions for the relative correlation lengths as in Ref. [1].  It was found that the $\chi^2$ values for the fits could be improved somewhat, especially for the high and low temperature scans, but at the cost of an arbitrary assumption about the relative behavior of the transverse and longitudinal lengths.  Importantly, this had neither a qualitative nor a quantitative effect on our overall conclusions so we present only the results with the thermal perpendicular correlation length and the fourth order thermal correction coefficient $c$ determined by the pure system relative behavior .

The temperature dependence of the peak position $q_o = 2\pi/d$, where $d$ is the smectic layer thickness, is shown in Fig. 10 for both global and free fits to several [8:6]OCB+aerosil samples with $\rho_s = 0.025$. In the SmA phase region, where the scattering peak is large and narrow, the uncertainty in the $q_o$ peak values is very small and the error bars are smaller than the size of the symbols plotted in the figure. In this region there is excellent agreement between $q_o$ values from global and free fits. Although Kortan *et al* [14] do not report any temperature dependence for $q_o$, our range of $q_o$ values in the SmA phase agrees well with the average values reported for the pure [8:6]OCB mixtures, as shown in Fig. 10. In the N and RN phase regions, the pretransitional smectic fluctuations yield weak x-ray scattering intensities. This in turn means that there are moderately large uncertainties in the $I(q)$ peak position for $T > T_1*$ and $T < T_2*$. This is evident from the error bars given in Fig. 10. Note that the $q_o$ values in the low-temperature RN region are significantly different for global and free fits, whereas the high-temperature nematic phase $q_o$ values are almost independent of the treatment of the background.

Although the large error bars for the $q_o(T)$ data in the N and RN phases make any detailed analysis of these data very difficult, some general comments can be made with confidence. The linear thermal expansion coefficient $\alpha_\parallel = (1/d)(\partial d/\partial T)$ is positive in the N phase and in the SmA phase just below $T_1*$ for both global and free fits. Indeed, $\alpha_\parallel$ exhibits a peak at $T_1*$, as expected from detailed studies of the critical linear thermal expansion observed near N-SmA transitions [20]. However, the behavior of $\alpha_\parallel$ near $T_2*$ is not well determined in our experiment since the global and free fits give very different $q_o(T)$ behavior in the RN phase. The $q_o(T)$ or $\alpha_\parallel$ behavior below $T_2$* from global fits is physically unlikely. In contrast, the free fits yield $\alpha_\parallel$ that is negative in the RN phase (like in the SmA phase just above $T_2$*), and $\alpha_\parallel$ exhibits a dip at $T_2$*. The volume thermal expansion coefficient $\alpha_V = (1/V)(\partial V / \partial T)$, as measured in a very high resolution dilatometric study of a [8:6]OCB mixture with $y = 0.406$, is positive at all temperatures [21]. In this [8:6]OCB mixture, $\alpha_V$ exhibits a positive peak at $T_2$* and a much smaller positive peak at $T_1$*. Thus the smectic layer thickness associated with fluctuations is similar for the N and RN phases, but the lateral packing of molecules in such "cybotactic" regions differs. The in-plane packing is substantially denser for the RN phase than for the N phase. This implies better orientational packing in the RN phase, which is consistent with the fact that the orientational order parameter $S$ increases monotonically on cooling and is substantially larger in the RN phase than in either the SmA or the N phase [22]. See Ref. 20 for further details of critical linear thermal expansion at nematic – smectic-A transitions.

## IV. DISCUSSION AND CONCLUSIONS

A high-resolution x-ray scattering study of [8:6]OCB+aerosil gels shows that the silica aerosil dispersion which forms a random gel network in the [8:6]OCB LC mixture also changes the phase transition characteristics of the N-SmA and SmA-RN transitions, although the fundamental N-SmA-RN reentrant sequence is still observed.

The effective transition temperatures $T_1$*(N-SmA) and $T_2$*(SmA-RN) are shown in Fig. 11 as functions of the [8:6]OCB composition variable $y$ for various sil concentrations $\rho_S$. Data on the 8OCB+aerosil system [3, 24] are also included. For $y \leq 0.33$, the $T$* values for the [8:6]OCB+aerosil gels agree quite well with those for pure [8:6]OCB mixtures. However, near the "nose", the SmA phase is systematically stabilized relative to the nematic (N or RN) phase by the addition of aerosil. In Ref. 14, no SmA phase was observed in pure [8:6]OCB mixtures when $y > y_o = 0.427$. Here, however, we find that a SmA phase is clearly established for a [8:6]OCB+aerosil sample with $y = 0.43$ and is even weakly seen for a sample with $y = 0.443$, but not seen for $y \geq 0.46$; see Table I. Since the SmA phase is weakly ordered in reentrant LC+aerosil samples, this suggests that the aerosil destabilizes the nematic phases more than it does the SmA phase in order to account for this stabilization of the SmA phase at slightly higher concentrations. It should be kept in mind that the free energy difference between SmA and N is extremely small near $y_o$ and thus even tiny changes in $G$ can cause visible shifts in the phase boundary. As a note of caution, we should also add that since in both Kortan et al.[14] and here the liquid crystal materials were used without further purification there could also be subtle effects due to differences in the purity and thus the effective concentrations.

There were two major experimental difficulties in this study. First, the temperature range that must be scanned was almost twice as large as that in experiments involving only a single N-SmA transition. Second, the time for the x-ray exposure of the samples was limited to a total of two hours, so that any possible x-ray damage would be avoided. Because of these restrictions, the density of temperature values at which scattering data were acquired was insufficient to allow reliable critical exponent analysis. In previous studies of LC+aerosil systems, the variation of $a_2(T)$ was describe by the power law $a_2 \sim \left| T - T^* \right|^{2\beta}$ since $a_2$ is proportional to the square of the smectic order parameter.  Such an analysis is very difficult here due to the sparsity of data points near $T_1$*and $T_2$* and the limited maximum temperature range over which any simple power-law variation could possibly hold.  Due to the crossover from  growth in the SmA order on cooling from $T_1$* down to $T_m$ and the subsequent decrease in SmA order on cooling from $T_m$ to $T_2$*, one cannot include in the power law fits any $a_2$ data near $T_m$.  Attempts to extract $\beta$ exponents from our data for samples with high sil concentrations yielded scattered $\beta$ values, mostly lying in the range 0.20 to 0.30. Only the $\beta$ values for $\rho_s = 0.025$   samples are listed in Table 1. Such values can be compared with that obtained for pure 8OCB using hyperscaling: $\beta = (2 - \alpha - \gamma)/2 = 0.24$ [7], but no systematic trends for $\beta$ with $y$ or $\rho_s$ can be established.

Although accurate values for the critical exponent $\beta$ cannot be obtained, the temperature behavior of the static amplitude $a_2(T)$ , shown in Fig. 3, was found to be a sensitive indicator of smectic ordering, and the qualitative behavior of  $a_2$ near $T_1$*and $T_2$* was consistent with previous studies [1-4].  Also the values of the $a_2(T)$ fitting parameter obtained from the scattering profile were quite insensitive to the choice of background or to variations in the ways the thermal parameters $\xi_\perp$  and $c$ were handled. Thus, the effective transition temperatures given in Table I were obtained from the $a_2(T)$ variation.  Alternatively, one could infer $T_1$*and $T_2$* values from the peak positions for the thermal correlation length $\xi_\parallel$ and thermal amplitude $\sigma_1$, but we find that these parameters are somewhat more sensitive to the fitting model than $a_2(T)$.

Different theoretical models have been used to describe the reentrant behavior of pure [8:6]OCB LCs mixtures [17-19].  These models all implicitly assume that the smectic order would be anomalously weak in reentrant systems.  Using the temperature behavior of the static structure amplitude $a_2(T)$ , we show that the short-range static smectic order in [8:6]OCB+aerosol gels is indeed weaker than the similar short-range static smectic order seen in 4O.8+aerosol gels [1].  Although this comparison is based on two LC+aerosol gel systems, it supports our conclusion that the maximum smectic order in the pure [8:6]OCB  LC mixture is weaker than that in pure 4O.8.

One uncertain aspect of this study is the temperature dependence of the background scattering (see Fig. 9).  Fortunately, detailed analyses have shown that changes from a global to a free background did not significantly alter any of the important fitting parameters in the temperature range from 300 K to 350 K.  The extent of any such small variations in the fitting parameters $\xi_\parallel$, $\sigma_1$, $\xi_{\parallel 2}$, and $a_2$ is indicated by the error bars given in Figs. 3 - 8 and in Table I. Below ~300 K, where $a_2$ becomes small or zero, changes in the background parameters (such as those shown in Fig. 9) affect mostly the value of  $q_o$, and to a lesser extent those of $\xi_\parallel$ and $\sigma_1$.  It was found that the latter two temperature-dependent fitting parameters had slightly smaller values for the free background fits than for global fits with temperature-independent $b_P$ and $b_c$ values.

As is evident from Fig. 10, the best fit values for $q_o$ in the N and RN phases are qualitatively similar to each other for free fits, whereas the RN $q_o(T)$ behavior is complicated and physically unattractive for global fits. There is, perhaps, a way to reconcile the dilemma of having either a T-independent background and an odd $q_o(T)$ RN behavior (global fits) or a T-dependent background and a more reasonable $q_o(T)$ RN behavior (free fits).  If one takes the background for the RN region to be independent of T (as might be expected physically) but adds a broad, weak, T-independent Lorentzian at $q_1$ $\approx 0.3$ Å$^{-1}$ in the RN region and assumes that the ratio $\xi_\perp / \xi_\parallel$ is a constant independent of $\xi_\parallel$ and hence independent of T, then the only fitting parameter that changes appreciably from the global values is $q_o(T)$ in the RN phase. The new $q_o(T)$ variation looks much like that shown in Fig. 10(*b*). The broad feature at $q_1$ would be explained as due to weak SmA$_1$ fluctuations in the RN phase just below the SmA$_d$-RN transition [25], and a T-independent $\xi_\perp / \xi_\parallel$ ratio is what one expects from isotropic scaling. Unfortunately, the present data below $T_2*$ are too sparse to test such a model.

In summary, we have carried out a detailed study of reentrant nematic behavior in 8OCB-6OCB mixtures with the liquid crystal material embedded in dilute aerosil gels. The aerosil gel exerts weak random fields which couple linearly to both the nematic and the SmA order parameters. We find that although the reentrant behavior is preserved, the SmA phase long range order is destroyed by the gel and replaced by static short range order. Even though the random field destroys the smectic long range order, the formation and subsequent destruction of smectic layers with decreasing temperature are readily observable and, in particular, one can observe clearly the diverging smectic correlation length and susceptibility at both transitions. These are not accessible in the aerosil-free systems because in those cases the smectic-A phase is a Landau-Peierls algebraic decay state.

## V. Acknowledgments

The work at the University of Toronto was supported by the Natural Science and Engineering Research Council of Canada and the work at Lawrence Berkeley Laboratory is supported by the Office of Basic Energy Sciences, U.S. Department of Energy under contract number DE-AC03-76SF00098. The x-ray scattering experiments were conducted at the National Synchrotron Light Source, Brookhaven National Laboratory, supported by US Department of Energy under Contract No. DE-AC-02-98CH10886, and at SSRL, a national user facility operated by Stanford University on behalf of DOE/BES.

**ρ_s = 0.025, L_r = 2700**

| | $y=0.33$ | $y=0.405$ | $y=0.42$ | $y=0.43$ | $y=0.46,\ 0.50$ |
|---|---|---|---|---|---|
| $\xi\|_2$ | $10000\pm880$ | $8200\pm600$ | $7300\pm500$ | $3700\pm250$ | $\rho_s\ \ 0.025$ |
| $\beta$ | $0.26\pm.01$, $0.24\pm.01$ | $0.25\pm.01$, $0.25\pm.03$ | $0.22\pm.03$, $0.17\pm.05$ | $0.36\pm.05$, $0.25\pm.10$ | $y\ \ 0.46$ |
| $a_2/b_p$ | $1.42\pm05$ | $0.76\pm.01$ | $0.74\pm.02$ | $0.22\pm03$ | $\sigma_t/b_p\ \ 83\pm9$ |
| $T^*$ | $325\pm5$, $294\pm.5$ | $318\pm1$, $302.5\pm.5$ | $318.2\pm6$, $302.6\pm.4$ | $317.7\pm.5$, $304\pm1$ | |

**ρ_s = 0.062, L_r = 1075**

| | $y=0.33$ | $y=0.405$ | $y=0.42$ | $y=0.443$ | $y=0.46,\ 0.50$ |
|---|---|---|---|---|---|
| $\xi\|_2$ | $3400\pm270$ | $1750\pm200$ | $3300\pm200$ | $2200\pm210$ | $\rho_s\ \ 0.061$ |
| $a_2/b_p$ | $0.34\pm.01$ | $0.11\pm.01$ | $0.28\pm.03$ | $0.043\pm.7$ | $y\ \ 0.46$ |
| $T^*$ | $324.5\pm0.5$, $297.4\pm0.5$ | $318.4\pm.8$, $303.5\pm.5$ | $320.3\pm.5$, $301.1\pm1$ | $315.7\pm.8$, $306\pm1$ | $\sigma_t/b_p\ \ 13.5\pm1$ |

**ρ_s = 0.15, L_r = 445**

| | $y=0.33$ | $y=0.405$ | $y=0.42$ | $y=0.43$ | $y=0.46,\ 0.50$ |
|---|---|---|---|---|---|
| | | | | $\boldsymbol{\rho_s}\|_2\ \ 0.117$ | |
| $\xi\|_2$ | $930\pm100$ | $730\pm95$ | $1000\pm85$ | $640\pm70$ | $\rho_s\ \ 0.025$ |
| $a_2/b_p$ | $0.063\pm.007$ | $0.068\pm.002$ | $0.093\pm.004$ | $0.018\pm.002$ | $y\ \ 0.5$ |
| $T^*$ | $324.5\pm1$, $297\pm1$ | $321.4\pm.5$, $302\pm1$ | $318\pm1$, $302.4\pm.8$ | $317.1\pm.7$, $305\pm3$ | $\sigma_t/b_p\ \ 13.3\pm.5$ |

Table I.  Parameters obtained from global fitting analysis of [8:6]OCB+aerosil samples. Listed are the gel densities $\rho_S$ in g of $SiO_2/cm^3$ of LC and the mean gel network pore sizes $l_o$ in Å [2];  the global ( $T$-independent) parallel correlation lengths for the static term of the structure factor $\xi_{\parallel 2}$ ; the maximum static amplitude $a_2$ normalized by the Porod amplitude $b_P$ in arbitrary (but common) units, and the transition temperatures $T_1^*$ (N-SmA) and $T_2^*$ (SmA-RN). The critical order parameter exponents $\beta$  are given only for the $\rho_S = 0.025$ gels (upper value for the N-SmA transition).  Note that no SmA phase was observed for $y$=0.46 and $y = 0.50$ concentrations.  For those concentrations, normalized thermal amplitudes $\sigma_1 / b_P$ (thermal amplitude of the structure factor over Porod amplitude) are given.

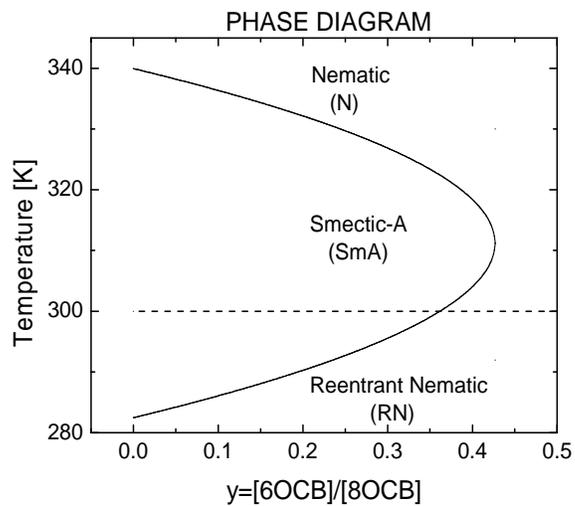

FIG.1. Phase diagram for [8:6]OCB liquid crystal mixtures [14]. The SmA-RN wing is difficult to study since both pure 8OCB and the mixtures freeze into a 3D-crystal phase at ~300 K (shown by the horizontal dashed line).

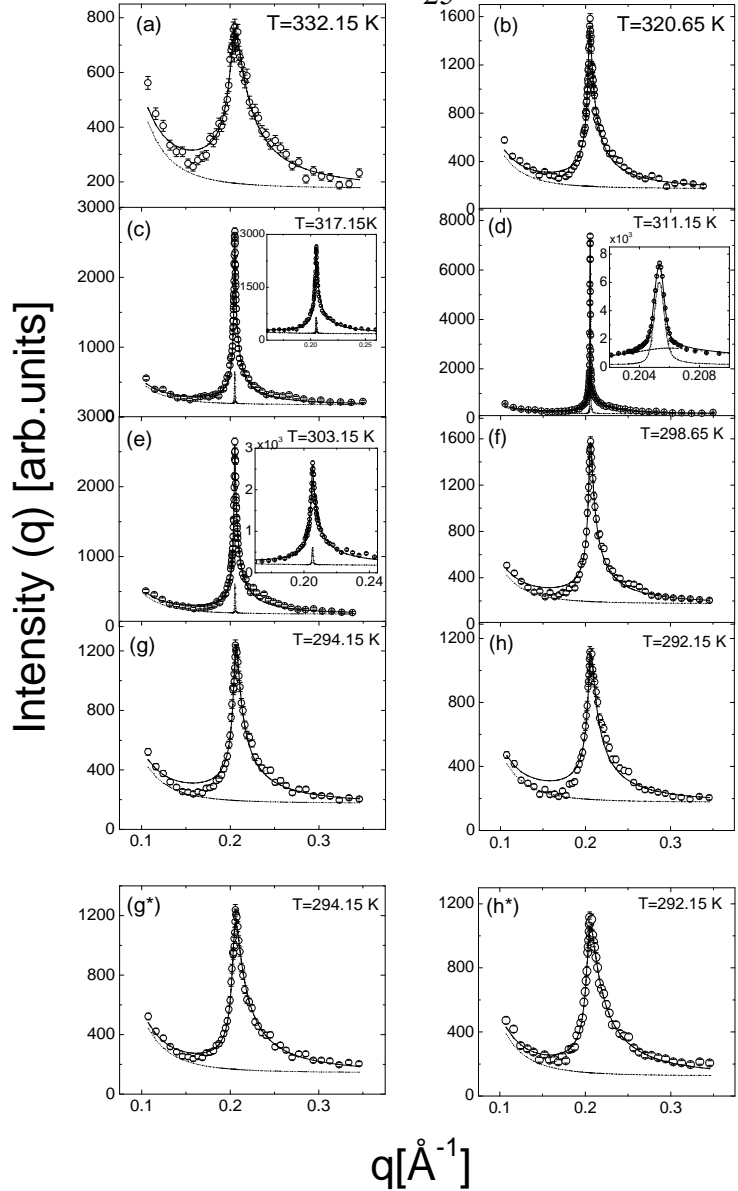

FIG. 2. Scattering intensity *I(q)* for a [8:6]OCB+aerosil sample with y = 0.405 and $\rho_S$ = 0.025 g cm$^{-3}$. The solid line in these panels represents the total structure factor, Eq. (1) plus the background. The thermal, Eq. (2) and static, Eq. (3) structure factors plus the background are shown by dashed and dotted lines, respectively. For panels a and b (high temperature region) the data are modeled using the thermal part of the structure factor only. The static structure factor starts to contribute to the profile in panel c, reaches its maximum in panel d, decreases through out panel e and totally diminishes in panel f while the remaining data is modeled using the thermal structure factor, like in the high temperature region, in panels f, g and h. In the global fitting analysis shown in panels *a - h*, the background parameters are taken to be independent of temperature. Panels *g\** and *h\** show the results of free fits with a temperature-dependent background. For panels *c-e*, the sample is in the SmA phase.

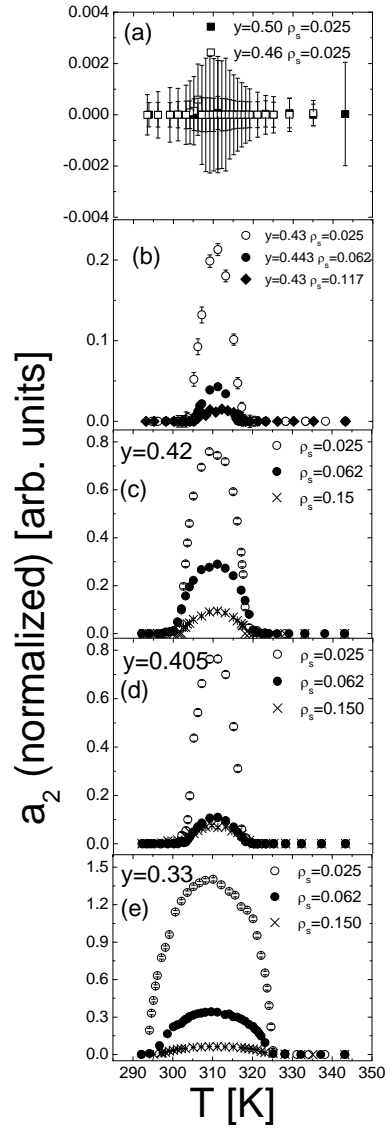

FIG. 3. Normalized static amplitude $a_2/b_p$ versus temperature obtained from global fits to $I(q)$. The normalization was achieved with the $T$-independent Porod background amplitude $b_p$. These plots show the effect of the sil density $\rho_S$ for the specified concentrations $y = [6OCB]/[8OCB]$ of [8:6]OCB+aerosil gel samples.

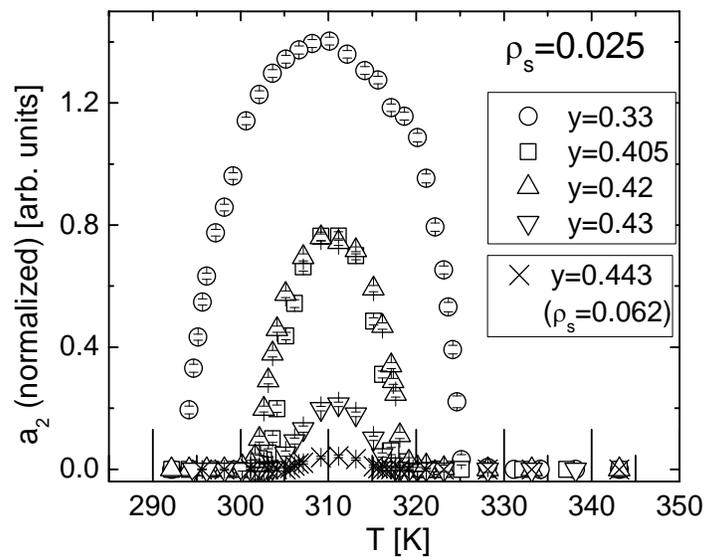

FIG. 4. The systematic decrease in normalized static amplitude, $a_2/b_P$, in the SmA phase is shown as a function of concentration $y$ for $\rho_S = 0.025$ gel samples. These parameters are a subset of the global fit parameters shown in Fig. 3. In the case of $y = 0.443$, only $\rho_S = 0.062$ data are available.

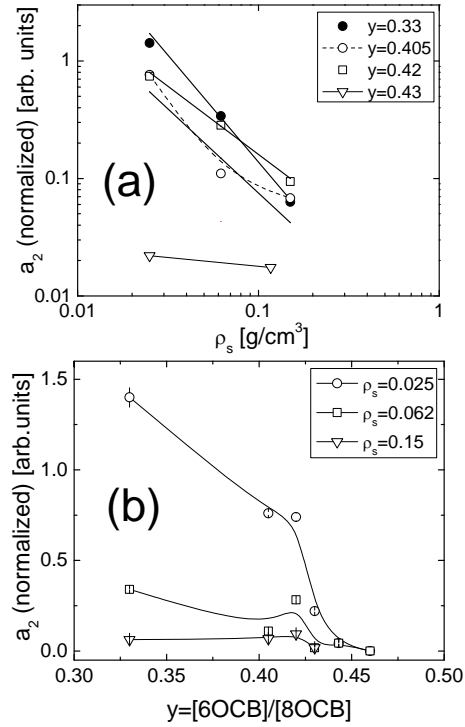

FIG. 5. Normalized maximum static amplitude $a_2/b_P$ versus sil density $\rho_S$ and versus LC concentration $y$. The maximum $a_2$ static amplitude values were obtained at $T=T_m$. The solid lines in panel $a$ represent the results of power-law fits with $a_2/b_P \sim \rho_S^{-x}$ with $x = 1.8(1), 1.4(1), 1.1(1), 0.3(1)$ for $y = 0.33$, $y = 0.405$, $y = 0.42$ and $y=0.43$, respectively. The lines in panel $b$ are guides for the eye.

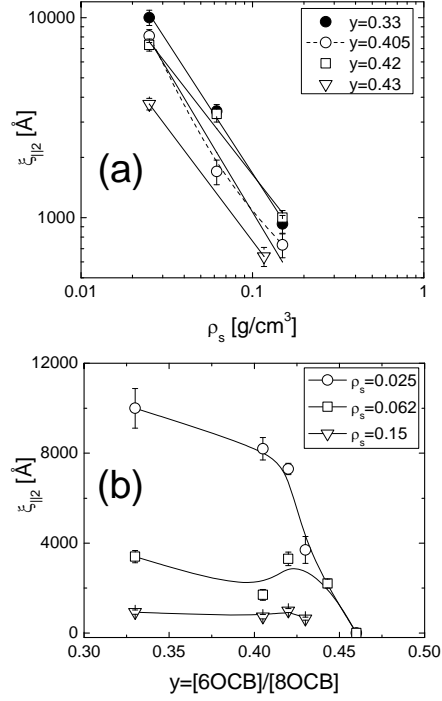

FIG. 6. Static correlation length $\xi_{\parallel 2}$ in Å versus sil density $\rho_s$ and versus LC

concentration y. The solid lines in panel *a* represent the results of power-law fits, $\xi_{\parallel 2} \sim \rho_S^{-z}$ with $z = 1.3(1)$, $1.4(1)$, $1.1(1)$ and $1.1(1)$ for $y = 0.33$, $0.405$, $0.42$ and $0.43$, respectively. The lines in panel *b* are guides for the eye.

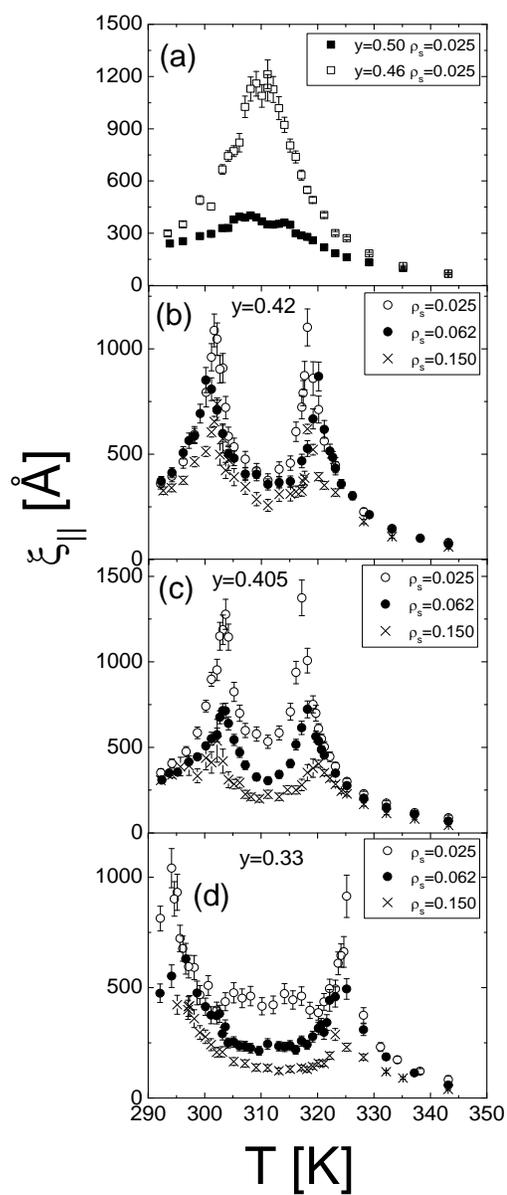

FIG. 7. Thermal parallel correlation length $\xi_{\parallel}$ versus temperature for several

[8:6]OCB+aerosil samples, as obtained from global fits.

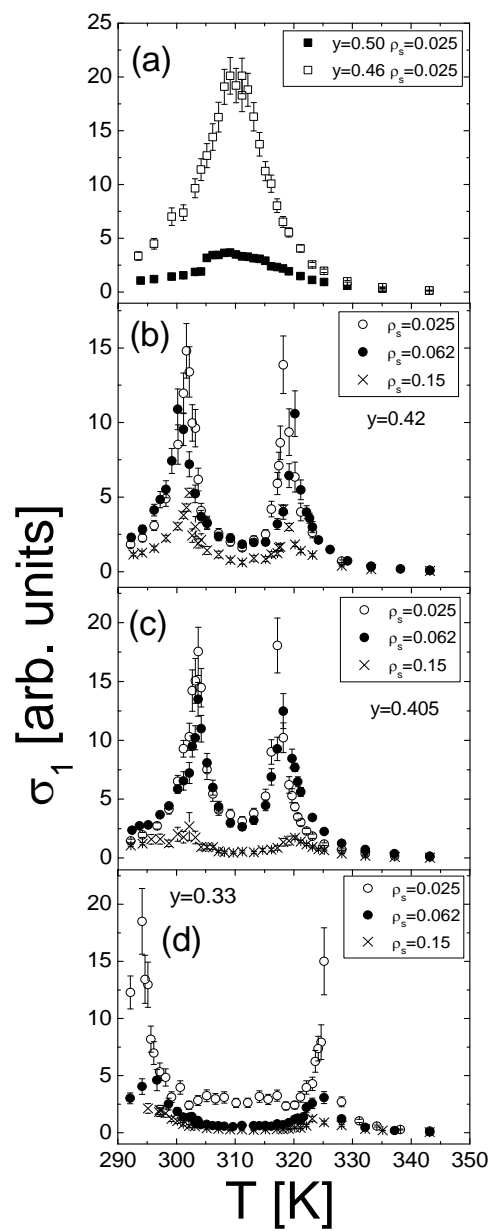

FIG 8. Thermal amplitude $\sigma_1$ versus temperature for several [8:6]OCB+aerosil samples, as obtained from global fits.

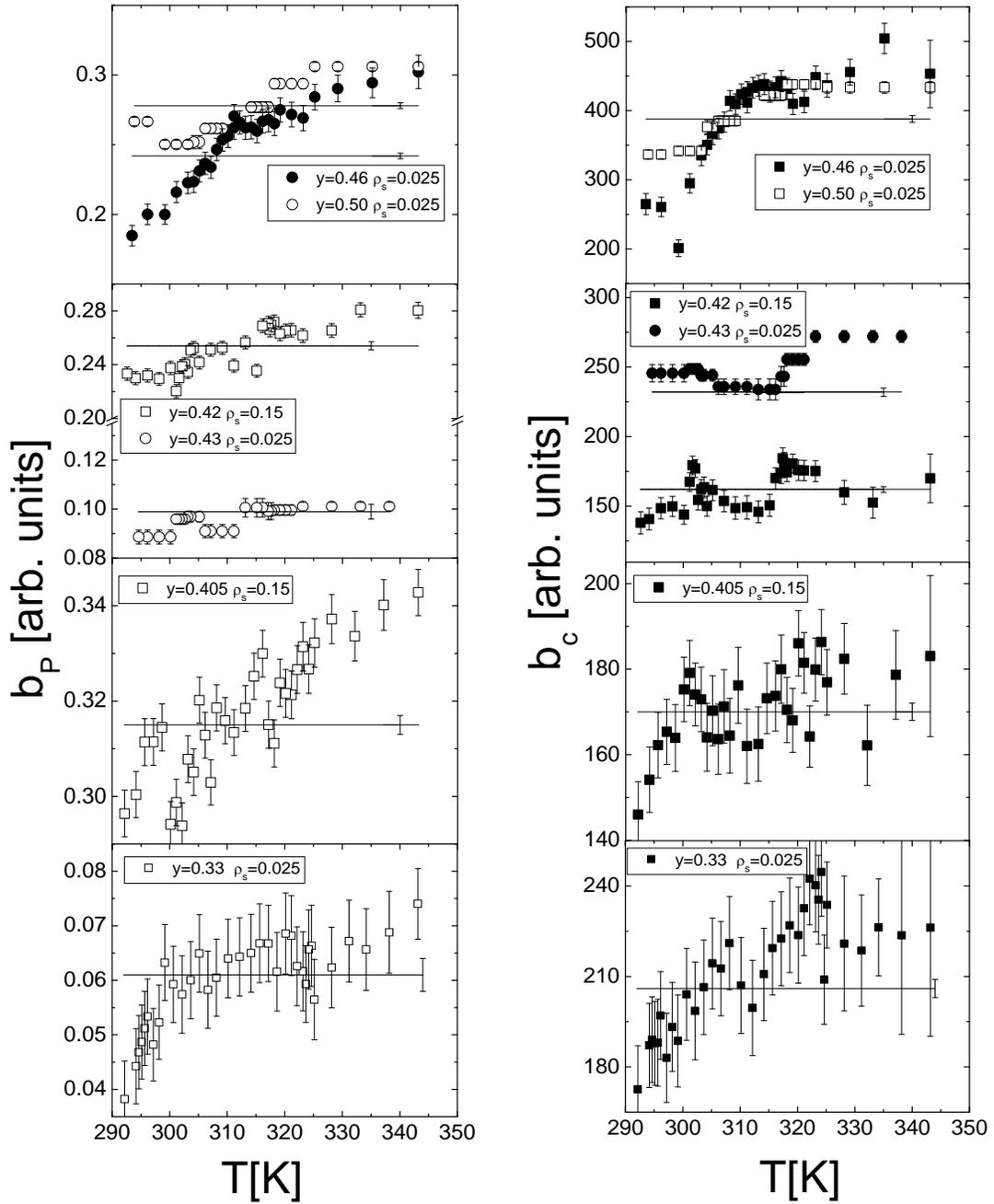

FIG. 9. Fitting parameters for the background $b_c + b_P/q^4$. The horizontal lines show the T-independent values obtained from global fits; the points are obtained from free fits.

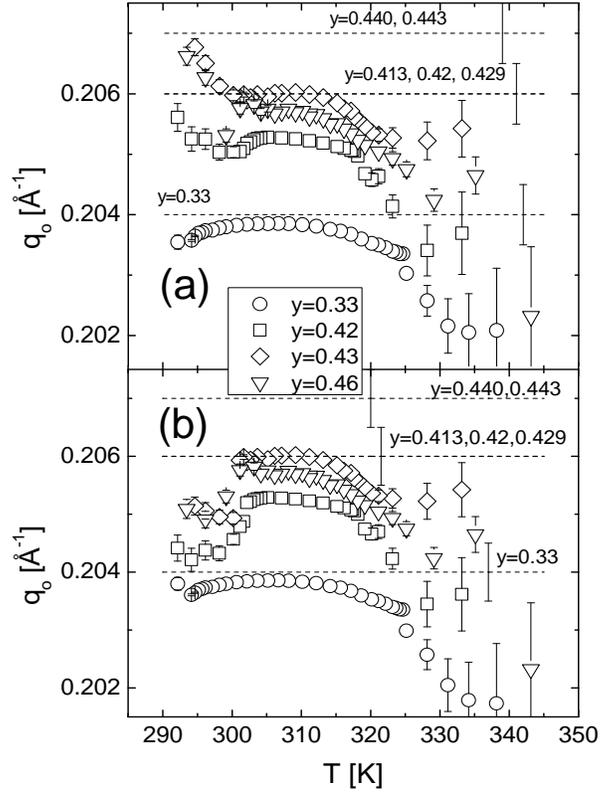

FIG. 10. Peak position $q_o$ versus temperature for [8:6]OCB+ aerosil samples with $\rho_S = 0.025$ obtained from global fits (panel $a$) and free fits (panel $b$). Note that the sample with $y = 0.46 > y_o$ does not exhibit a SmA phase. The $q_o$ values in the SmA phase are essentially identical for the two fitting models, and there is fairly good agreement also in the N phase. However, $q_o(T)$ in the RN phase differs significantly for global and free fits (see text). The $q_o$ values for pure LC mixtures [14] with $y = 0.33$, $y = 0.413$-$0.429$, and $y = 0.440$-$0.443$ are indicated by dashed horizontal lines.

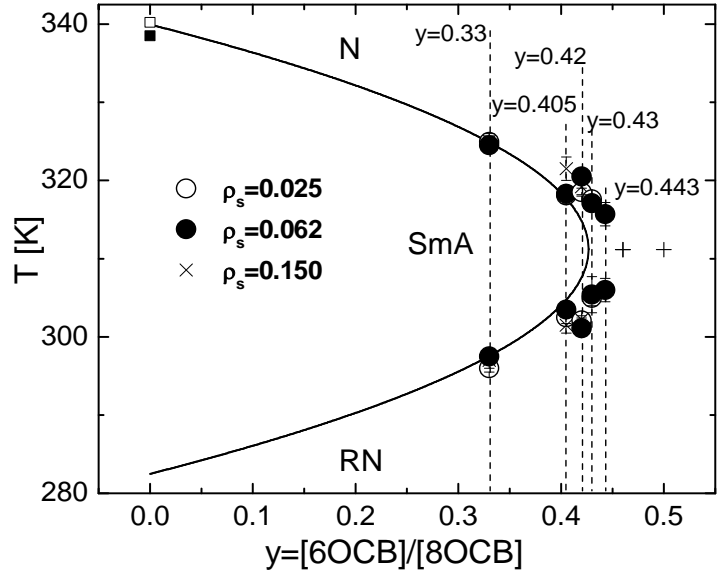

FIG. 11. Phase diagram for [6:8]OCB+aerosil gels. The solid line shows the parabolic phase boundary for the pure LC mixtures [14, 21, 23]. The dashed vertical lines indicate the different sample compositions studied here. The plus signs show the position of the peaks in $\xi_\parallel$ and $\sigma_1$ for samples with $y = 0.46$ and $y = 0.50$, for which no SmA phases was detected. For 8OCB+aerosil (i.e., $y = 0$), data were taken from Ref. 24 at $\rho_S$ values of 0.051 (open square) and 0.105 (closed square).